\documentclass[journal,twoside, a4paper]{IEEEtran}
\IEEEoverridecommandlockouts
\usepackage{cite}
\usepackage{amsmath,amssymb,amsfonts}
\usepackage{graphicx}
\usepackage{textcomp}
\usepackage{xcolor}
\usepackage{hyperref}
\usepackage[ruled,vlined,noresetcount]{algorithm2e} 

\usepackage{algpseudocode}

\usepackage{amsmath}
\usepackage{booktabs} 
\usepackage{tikz}
\usetikzlibrary{shapes}
\usetikzlibrary{arrows}

\usepackage{subcaption}
\usepackage{pgfplots}
\pgfplotsset{width=10cm,compat=1.15}

\def\BibTeX{{\rm B\kern-.05em{\sc i\kern-.025em b}\kern-.08em
    T\kern-.1667em\lower.7ex\hbox{E}\kern-.125emX}}
\usepackage[a4paper, total={7in, 9in}]{geometry}
\begin{document}

\title{Complexity of Post-Quantum Cryptography in Embedded Systems and Its Optimization Strategies}

% title2: Complexities in Post-Quantum Cryptography: Hardware Challenges and Potential Solutions

\author{\IEEEauthorblockN{Omar Alnaseri\IEEEauthorrefmark{1}, Yassine Himeur\IEEEauthorrefmark{5}, Shadi Atalla\IEEEauthorrefmark{5} and Wathiq Mansoor\IEEEauthorrefmark{5}
% Montgomery Scott\IEEEauthorrefmark{3}, and
% Eldon Tyrell\IEEEauthorrefmark{4},~\IEEEmembership{Fellow,~IEEE}
}
\\

\IEEEauthorblockA{\IEEEauthorrefmark{1}Department of Electrical Engineering, DHBW University, Ravensburg, Germany }\\
\IEEEauthorblockA{\IEEEauthorrefmark{5}College of Engineering and Information Technology, University of Dubai, Dubai, United Arab Emirates}
}

\maketitle

\pagenumbering{gobble}

\begin{abstract}
With the rapid advancements in quantum computing, traditional cryptographic schemes like Rivest-Shamir-Adleman (RSA) and elliptic curve cryptography (ECC) are becoming vulnerable, necessitating the development of quantum-resistant algorithms. The National Institute of Standards and Technology (NIST) has initiated a standardization process for PQC algorithms, and several candidates, including CRYSTALS-Kyber and McEliece, have reached the final stages. This paper first provides a comprehensive analysis of the hardware complexity of post-quantum cryptography (PQC) in embedded systems, categorizing PQC algorithms into families based on their underlying mathematical problems: lattice-based, code-based, hash-based and multivariate / isogeny-based schemes. Each family presents distinct computational, memory, and energy profiles, making them suitable for different use cases. To address these challenges, this paper discusses optimization strategies such as pipelining, parallelization, and high-level synthesis (HLS), which can improve the performance and energy efficiency of PQC implementations. Finally, a detailed complexity analysis of CRYSTALS-Kyber and McEliece, comparing their key generation, encryption, and decryption processes in terms of computational complexity, has been conducted. 
\end{abstract}


\section{Introduction}
Post-quantum cryptography (PQC) aims to develop cryptographic algorithms that are secure against attacks from quantum computers. With rapid advancements in quantum computing, traditional cryptographic schemes such as Rivest-Shamir-Adleman (RSA) and elliptic curve cryptography (ECC) are becoming vulnerable. PQC algorithms are broadly categorized into families based on their underlying mathematical problems, each posing distinct challenges for implementation in embedded systems. These families include lattice-based, code-based, hash-based, and multivariate/isogeny-based schemes, each with unique computational, memory, and energy profiles.

\begin{table*}[h]
    \centering
    \caption{Comparison of Cryptographic Families}
    \label{tab:cryptographic_families}
    % \resizebox{\textwidth}{!}{%
    \begin{tabular}{lcccc}
        \toprule
        \textbf{Family} & \textbf{Computational Complexity} & \textbf{Memory Footprint} & \textbf{Energy Efficiency} & \textbf{Implementation Flexibility} \\
        \midrule
        Lattice-Based & Medium-High (NTT) & Medium (1–3 kB RAM) & Medium & High (scalable parameters) \\
        Code-Based & Low & Very High ($>$1 MB) & Low & Low (fixed key sizes) \\
        Hash-Based & Low & Low ($<$2 kB RAM) & Very High & Medium (state management) \\
        Multivariate & Very High & High ($~$100 kB) & Very Low & Low \\
        Isogeny-Based & Very High & Medium ($~$10 kB) & Very Low & Low \\
        \bottomrule
    \end{tabular}
    % }
\end{table*}

Lattice-based algorithms derive security from hard problems like the shortest vector problem (SVP) or learning with errors (LWE). Schemes such as Kyber (key encapsulation) and CRYSTALS-Dilithium (digital signatures) rely on polynomial arithmetic and the number theoretic transform (NTT) for efficient polynomial multiplication. While NTT speeds up computations, it introduces hardware overhead due to modular arithmetic and parallel operations. Ducas et al. \cite{Ducas2017CRYSTALSDilithiumAS} showed NTT optimizations reduce latency by up to 40\% on embedded platforms. However, lattice-based algorithms still require moderate memory (1-3 kB RAM for keys) and scalable parameters, making them demanding for limited devices. Kannwischer et al. \cite{kannwischer2024pqm4} demonstrated Kyber on ARM Cortex M4, using less than 10 kB RAM, although it is less energy efficient than hash-based schemes.

Code-based algorithms, such as McEliece and BIKE, rely on decoding random linear codes. They are computationally lightweight, using matrix multiplications and sparse linear algebra, but suffer from large key sizes (often over 1 MB). Chou et al. \cite{chen2021classic} reduced McEliece keys by 50\% on STM32 microcontrollers, but compressed keys still overload flash storage, limiting their use in memory-constrained devices. 

Hash-based algorithms, such as SPHINCS+ and XMSS \cite{Buchmann2011XMSSA}, use cryptographic hash functions (e.g. SHA-3 or SHAKE-256). They are computationally simple, relying on iterative hashing, and are lightweight and parallelizable. Bernstein et al. \cite{bernstein2019sphincs} showed SPHINCS+ runs on 8-bit AVR microcontrollers with just 2 kB RAM, making it suitable for energy-constrained IoT devices. However, stateless schemes like SPHINCS+ produce large signatures (up to 41 KB), while stateful schemes (e.g., XMSS) require nonvolatile memory (NVM) for state tracking. Bernstein et al. \cite{Bernstein2017gimli} noted that parallelizing hash chains improves throughput but increases area overhead, limiting cost-sensitive applications.

Multivariate and isogeny-based algorithms are niche categories with limited embedded adoption. Multivariate schemes like Rainbow involve solving nonlinear polynomial equations, leading to computationally intensive operations and large keys (often over 100 kB). Isogeny-based schemes, such as SIKE \cite{azarderakhsh2017supersingular}, use complex elliptic curve isogeny computations that are sequential and hard to accelerate. Koziel et al. \cite{koziel2016post} showed that SIDH FPGA implementations require significant area and power, even with optimizations. Table~\ref{tab:cryptographic_families} compares cryptographic families in computational complexity, memory footprint, energy efficiency, and flexibility.

The National Institute of Standards and Technology (NIST) has narrowed the candidates to a few finalists,including CRYSTALS-Kyber and McEliece  \cite{bandaru2024evaluation}. Nonetheless, embedded systems, characterized by their limited computational power and memory, present unique challenges for PQC implementation. Achieving successful deployment of these candidates on embedded systems necessitates a comprehensive analysis of the hardware complexity associated with these PQC algorithms. Thus, this paper conducts an exhaustive examination of the hardware complexity involved in deploying PQC within such systems, with a particular emphasis on CRYSTALS-Kyber and McEliece. This paper makes several key contributions to the field of PQC in embedded systems: 
\begin{itemize}
    \item Comprehensive analysis: The paper provides a detailed analysis of the hardware complexity of PQC algorithms, categorizing them into families based on their underlying mathematical problems, and discussing their computational, memory, and energy profiles.
    \item Optimization strategies: The paper explores optimization strategies such as pipelining, parallelization, and high-level synthesis (HLS) to improve the performance and energy efficiency of PQC implementations in embedded systems.
    \item Complexity comparison: It offers a detailed complexity analysis of two leading PQC candidates, CRYSTALS-Kyber and McEliece, comparing their key generation, encryption, and decryption processes in terms of computational complexity, memory footprint, and energy efficiency.
    % \item Practical implications The analysis provides practical insights into the trade-offs between security, performance, and resource usage, helping researchers and practitioners select and optimize PQC algorithms for embedded systems.
\end{itemize}

% \section{Hardware Implementation Challenges}
% Embedded systems are widely used in various applications, from IoT devices to automotive systems. Their limited resources make them challenging environments for implementing complex cryptographic algorithms. The long life of embedded devices necessitates the adoption of PQC to ensure long-term security \cite{pursche2024sok}.

% \subsection{Computational Complexity}
% PQC algorithms often require more computational power due to larger keys and complex arithmetic operations. For example, lattice-based schemes involve matrix multiplications and polynomial arithmetic, which can be computationally intensive \cite{xie2020special}.

% \subsection{Memory Constraints}
% Embedded systems have limited memory, making it difficult to store large keys and intermediate results required by PQC algorithms. Code-based schemes like McEliece require significant memory for storing large matrices \cite{bandaru2024evaluation}.

% \subsection{Side-Channel and Fault Attacks}
% PQC implementations in hardware must be protected against side-channel and fault attacks. These attacks exploit physical characteristics of the hardware to extract secret information. Designing PQC hardware that is resistant to such attacks adds to the complexity \cite{xie2020special}.

\section{Optimization Strategies}
In this section, we propose optimization strategies that improve the performance of PQC algorithms in embedded systems by focusing on techniques that improve computational efficiency and resource management. We will explore methods such as pipelining, parallelization, and efficient use of hardware resources to maximize algorithmic throughput and reduce execution time.
\subsection{Pipelining and Parallelization}
% Pipelining and parallelization can be used to optimize the performance of PQC algorithms in embedded systems. 
% For example, parallelizing matrix multiplications in lattice-based schemes can significantly reduce computation time. 

Pipelining involves breaking down a complex task into a series of smaller tasks, each performed at a different stage of a pipeline \cite{Zhao2021ACA}. This allows for parallel processing, where multiple tasks can be executed simultaneously, reducing the overall processing time. Pipelining can be applied to various stages of the cryptographic process, such as key generation, signature generation, and signature verification \cite{Liu2024HighPerformanceHI}. For example, the CRYSTALS-Dilithium scheme uses a pipelined processing method to reduce both storage requirements and processing time \cite{Zhao2021ACA}. Similarly, the Picnic digital signature scheme uses a pipelined approach to optimize its hardware implementation, resulting in a significant reduction in clock cycle count and energy consumption \cite{Liu2024HighPerformanceHI}.

Parallelization, on the other hand, involves dividing a task into smaller subtasks that can be executed concurrently by multiple processing units \cite{Fritzmann2020RISQVTC}. This can be achieved using parallel architectures, such as multi-core processors or specialized accelerators. For instance, the RISQ-V architecture \cite{Fritzmann2020RISQVTC} integrates tightly coupled accelerators directly into the processing pipeline to speed up lattice-based cryptography. The accelerators include an arithmetic unit for vectorized modular arithmetic and NTT operations, a vectorized modular multiply accumulate unit, a Keccak accelerator for the pseudo-random bit generation, and a binomial sampling unit for the generation of binomially distributed samples.

Pipelining and parallelization can be combined to further improve performance. For example, the design of the "coding for energy reduction with multiple encryption techniques" (CERMET) architecture incorporates both pipelining and process parallelization to improve efficiency \cite{Woo2023CERMETCF}. The system operates fully pipelined, ensuring that no throughput is lost compared to a conventional cryptographic system, and maintains throughput despite additional data processing steps. However, pipelining and parallelization can also introduce additional complexity and overhead. For example, the polynomial factorization method used in parallel quantum signal processing can reduce the depth of the query by a factor $O(k)$, but it comes with an increased number of measurements $O(poly(d)2^{O(k)})$ \cite{Martyn2024ParallelQS}, where $k$ is the module rank. 
% Similarly, optimizing modular reduction in hardware for arbitrary static moduli can be computationally expensive and pose a performance bottleneck \cite{Mller2023AreaEM}.

\subsection{High-Level Synthesis (HLS)}
% HLS can be used to design accelerators for PQC algorithms. It allows for rapid prototyping and optimization of hardware designs, making it easier to explore the design space and find efficient implementations.
% It is a design automation technique that enables the creation of hardware accelerators from high-level descriptions of algorithms \cite{Liao2022AHS}. HLS can be used to optimize the implementation of PQC algorithms to improve performance, energy efficiency, and security.
% An HLS optimization strategy for PQC is the use of a hybrid HLS approach, which combines state-based HLS with performance-driven HLS. This approach enables the design and optimization of application-specific embedded systems with precise timing behaviors, which is critical for cryptographic applications. The hybrid HLS approach utilizes explicit and precise timing information within periodic state machine (PSM) models to effectively reduce energy consumption \cite{Liao2022AHS}. 
It automates design, creating hardware from algorithm descriptions. It optimizes post-quantum cryptography (PQC) for better performance, energy efficiency, and security. A hybrid HLS strategy combines state-based and performance-driven approaches, using periodic state machine models for precise timing and reduced energy use \cite{Liao2022AHS}. Another HLS optimization strategy is the use of a hierarchical post-route quality of results (QoR) prediction approach. This approach estimates latency and post-route resource usage from C/C++ programs and uses a graph construction method to represent the control and data flow graph of source code and the effects of HLS pragmas. The approach also uses a hierarchical graph neural network (GNN) training and prediction method to capture the impact of loop hierarchies \cite{Gao2024HierarchicalSQ}. However, HLS optimizations can also affect the security and reliability of cryptographic implementations. For example, HLS optimizations can compromise the properties of countermeasures implemented using HLS, such as masking and hiding countermeasures. Therefore, secure circuit designers should be careful when using an HLS flow to integrate SCA countermeasures \cite{Koufopoulou2022SecurityAR}.

\subsection{Algorithmic Optimizations}
Optimizing the algorithms themselves can also reduce the hardware complexity. For example, using more efficient mathematical techniques can help reduce computational and memory requirements. Difference optimization strategies can be employed to improve the performance of PQC algorithms, such as:

\subsubsection{Hybrid approach} combines quantum key distribution (QKD) with PQC for authentication purposes \cite{Yunakovsky2021TowardsSR}. This can be particularly useful for protecting highly loaded communications links at a distance, where intermediate nodes may not be necessary. Additionally, standardization processes, such as those led by the NIST, can help identify and standardize post-quantum algorithms for stateless digital signatures and key encapsulation mechanisms/public key encryption.

\subsubsection{Signature lifting} allows users who failed to migrate to PQC in time to still use pre-quantum signature schemes while protecting against quantum attacks \cite{Sattath2023ProtectingQP}. This can be achieved by lifting a deployed pre-quantum signature scheme satisfying a certain property to a post-quantum signature scheme that uses the same keys.

\subsubsection{Quantum approximations} can also be beneficial for optimizing PQC algorithms. For example, a quantum mean value approximation can be used to approximate the density of the lattice basis, which can be used to improve the performance of lattice-based cryptography \cite{Joseph2021QuantumMA}.

\subsubsection{Quantum binary field multiplication} can optimize PQC operations, which can achieve a Toffoli depth of one for any field size, making it more efficient for quantum cryptanalysis of ECC \cite{Jang2023QuantumBF}.

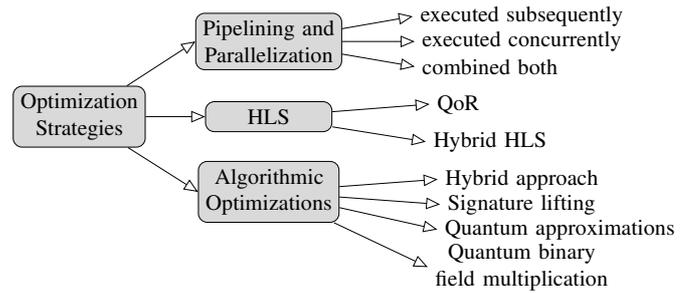
\begin{figure}[t!]
    \centering
    \resizebox{0.5\textwidth}{!}{%
\begin{tikzpicture}[
    node distance=1.2cm,
    % every node/.style={draw, rectangle, rounded corners, minimum width=3cm},
    every path/.style={draw, ->, >=open triangle 45},
    styleA/.style={draw, rectangle, rounded corners, minimum width=2cm},
    styleB/.style={draw=none}]

    % Nodes
    \node[styleA] (math) [align=center, fill=gray!30] {\baselineskip=12pt Optimization\\ Strategies};
	\node[styleA] (Lim_sc) [align=center, fill=gray!30, right of=math, node distance=3cm] {HLS};

    \node[styleB] (HLS_1) [align=center, right of=Lim_sc, yshift=-0.4cm, node distance=3.5cm]{Hybrid HLS};

    \node[styleB] (HLS_2) [align=center, right of=Lim_sc, yshift=0.2cm, node distance=3cm]{QoR};
   
	\node[styleA] (Sim_ass) [align=center, fill=gray!30, above of=Lim_sc] {Pipelining and \\ Parallelization};
    
    \node[styleB] (pip_1) [align=center, right of=Sim_ass, yshift=0.4cm, node distance=4cm]{executed subsequently};
    
    \node[styleB] (pip_2) [align=center, right of=Sim_ass, yshift=0cm, node distance=4cm]{executed concurrently};
       
    \node[styleB] (pip_3) [align=center, right of=Sim_ass, yshift=-0.4cm, node distance=3.5cm]{combined both};
   
    \node[styleA] (lac_ins) [align=center, fill=gray!30, below of=Lim_sc] {Algorithmic\\ Optimizations};
    
    \node[styleB] (Alg_1) [align=center, right of=lac_ins, yshift=0.2cm, node distance=4cm]{Hybrid approach};
    
    \node[styleB] (Alg_2) [align=center, right of=lac_ins, yshift=-0.2cm, node distance=4cm]{Signature lifting};
    
    \node[styleB] (Alg_3) [align=center, right of=lac_ins, yshift=-0.6cm, node distance=4.6cm]{Quantum approximations};
    
    \node[styleB] (Alg_4) [align=center, right of=lac_ins, yshift=-1.2cm, node distance=4cm]{Quantum binary \\field multiplication};

    % Connections
    \draw (math) to(Lim_sc);
    \draw (math) to(Sim_ass.west);
    \draw (math) to(lac_ins.west);
    \draw (Lim_sc) to(HLS_1.west);
    \draw (Lim_sc) to(HLS_2.west);
    \draw (lac_ins) to(Alg_1.west);
    \draw (lac_ins) to(Alg_2.west);
    \draw (lac_ins) to(Alg_3.west);
    \draw (lac_ins) to(Alg_4.west);
    \draw (Sim_ass) to(pip_1.west);
    \draw (Sim_ass) to(pip_2.west);
    \draw (Sim_ass) to(pip_3.west);
    
\end{tikzpicture}
}
    \caption{Optimization Strategies}
    \label{fig_opt_str}
\end{figure}

Algorithm \ref{algo1} summarizes the proposed process to optimize the implementation of PQC in embedded systems, specifically for the CRYSTALS-Kyber and McEliece schemes. It consists of key generation, encryption, and decryption. Kyber constructs a public key using a polynomial matrix, while McEliece employs a Goppa code with scrambling and permutation matrices. The encryption and decryption processes involve modular arithmetic and error correction. The algorithm integrates pipelining, high-level synthesis, modular reduction, and memory optimization to enhance efficiency while maintaining security in embedded systems.

\begin{algorithm}[t!]
    \SetAlgoLined
    % \color{blue}
    \caption{Optimized Post-Quantum Cryptography Implementation in Embedded Systems}
    \label{algo1}
    \KwIn{Security parameter $k$, Polynomial degree $n$, PQC scheme (Kyber or McEliece)}
    \KwOut{Optimized Key Generation, Encryption, and Decryption}
    
    \tcc{Step 1: Key Generation}
    \If{Scheme == Kyber}{
        Generate random matrix $A \in R^{k\times k}_q$\;
        Sample secret key vector $s \in R^k_q$ and error vector $e \in R^k_q$\;
        Compute public key: $t = A \cdot s + e$\;
    }
    \ElseIf{Scheme == McEliece}{
        Choose Goppa code parameters $(n, k, t)$\;
        Generate scrambling matrix $S \in \mathbb{F}_2^{k \times k}$ and permutation matrix $P \in \mathbb{F}_2^{n \times n}$\;
        Compute public key: $G' = S \cdot G \cdot P$\;
    }
    
    \tcc{Step 2: Encryption}
    \If{Scheme == Kyber}{
        Generate randomness $r \in R^k_q$ and error terms $e_1, e_2$\;
        Compute ciphertext: $u = A^T \cdot r + e_1$, \quad $v = t^T \cdot r + e_2 + \text{encode}(m)$\;
    }
    \ElseIf{Scheme == McEliece}{
        Compute codeword: $c = m \cdot G'$\;
        Generate random error vector $e \in \mathbb{F}_2^n$ with Hamming weight $\leq t$\;
        Compute ciphertext: $y = c + e$\;
    }
    
    \tcc{Step 3: Decryption}
    \If{Scheme == Kyber}{
        Recover message: $m = \text{decode}(v - s^T \cdot u)$\;
    }
    \ElseIf{Scheme == McEliece}{
        Apply permutation: $y' = y \cdot P^{-1}$\;
        Decode using Goppa decoding algorithm to recover $c'$\;
        Recover message: $m = c' \cdot G^{-1} \cdot S^{-1}$\;
    }
    
    \tcc{Optimization Strategies}
    \ForEach{Optimization technique in [Pipelining, HLS, Modular Reduction, Memory Optimization]}{
        Apply technique to relevant PQC operation\;
    }
    
    \Return Optimized Key Generation, Encryption, and Decryption\;
\end{algorithm}

\vskip-3mm

\section{Complexity Analysis}
\subsection{CRYSTALS-Kyber}
% CRYSTALS-KYBER is a lattice-based key encapsulation mechanism. It has been implemented in hardware with optimizations like pipelining and parallelization, achieving high throughput and low latency.
CRYSTALS-Kyber is a lattice-based cryptosystem that relies on the module learning with errors (Module-LWE) problem. The security of Kyber is based on the difficulty of solving the Module-LWE problem, even for quantum computers, making it a strong candidate for post-quantum cryptography. The Kyber is a key encapsulation mechanism (KEM) that is part of the "cryptographic suite for algebraic lattices" (CRYSTALS) suite, which is designed to be secure against quantum computers. It operates in the ring $R_q=Z_q[X]/(X^n+1)$, where $q$ is a prime modulus, $n$ is the degree of polynomial, typically a power of 2, and $X^n+1$ is the irreducible polynomial defining the ring \cite{Ghashghaei2024EnhancingTS}. To analyze the complexity, the operations primarily involve matrix-vector and vector-vector operations over polynomial rings. A breakdown of key operations of the kyber and their complexities is
% It involves generating a public key and a secret key, encapsulating a shared secret using the public key, and decapsulating it using the secret key. 

\begin{itemize}
    \item The key generation process involves first generate matrix $A \in R^{k\times k}_q$, where $k$ is the security parameter, typically 2, 3, or 4. Then generate a secret key vector $s\in R^k_q$ from a centered binomial distribution $\eta$. And finally generate public key $t=A\cdot s +e$, where $e\in R^k_1$ is a small error vector sampled from the same distribution $\eta$, which is a discrete Gaussian-like distribution. This operation has a complexity of $O(k^2 \cdot n)$, where $k$ is the module rank, i.e. 2, 3, or 4, and $n$ is the polynomial degree, e.g. 256. Therefore the FLOPs for this matrix-vector multiplication is $2k^2n$.
    \item The encryption process involves first generate randomness $r\in R^k_q$ from the distribution $\eta$, then compute ciphertext $(u,v)$, as $u=A^T\cdot r+e_1$, and $v=t^T\cdot r+e2+encode(m)$, where $e_1\in R^k_q$ and $e_2\in R_q$ are small error, and $encode(m)$ is the encoded message $m$.Each matrix-vector multiplication in encryption has a complexity of $O(k^2\cdot n)$, which results in a total of $2k^2n$ FLOPs.
    \item The decryption process involves first recovering the message $m$ by computing $v-s^T\cdot u$, which should be close to $encode(m)$ due to small errors. This is an operation of the inner product with a complexity of $O(k.n)$. This results in $2kn$ FLOPS.
\end{itemize}
Overall, Kyber is efficient in terms of key generation and decryption, with smaller key sizes and lower computational overhead, making it suitable for constrained environments.

\subsection{McEliece}
% The McEliece cryptosystem, a code-based scheme, has been implemented in hardware with optimizations focusing on reducing memory usage and improving computation speed. Techniques like quasi-cyclic codes have been used to reduce the size of the public key.
The McEliece cryptosystem is one of the earliest public-key cryptosystems, proposed by Robert McEliece in 1978 \cite{mceliece1978public}. It is based on error-correcting codes, specifically binary Goppa codes, and is considered a strong candidate for post-quantum cryptography because of its resistance to attacks by quantum computers. It relies on the hardness of decoding a random linear code, which is a well-known problem in coding theory. The key components are: (1) linear codes $C$ of length $n$ and dimension $k$ over a finite field $\mathbb{F}_q$ are a subspace $k$ of dimensions of $\mathbb{F}^n_q$. It can be represented by a generator matrix $G$ of size $k\times n$. And (2) Goppa codes, which is a specific class of linear codes with efficient decoding algorithms. Goppa codes are used in McEliece because they allow for efficient error correction. The security of McEliece relies on the fact that decoding a random linear code is a hard problem, even for quantum computers. To analyze the complexity of McEliece operations, it involves matrix multiplications, inversions, and error correction. A breakdown of its key operations and their complexities is provided below.

\begin{itemize}
    \item The key generation process involves first choosing a Goppa code $C$ with parameters $(n, k, t)$, which are the length, dimension of the codes, and error correction ability, respectively. Then generate a random scrambling matrix $S$ of $k\times k$ over $\mathbb{F}_2$, and generate a random permutation matrix $P$ of $n\times n$. Finally, compute the transformed generator matrix $G' = S\cdot G\cdot P$, so the public key is $(G',t)$, where $t$ is the error correction capability. This matrix-matrix multiplication has a complexity of $O(n^3)$, where $n$ is the length of the code. Thus, the number of FLOPs is $2n^3$. 
    \item The encryption process involves encrypting a message $m\in \mathbb{F}^k_2$ by computing the codeword $c=m\cdot G'$ and generating a random error vector $e\in \mathbb{F}^n_2$ with Hamming weight $wt(e)\le t$, and finally the ciphertext is calculated $y=c+e$. This matrix vector multiplication has a complexity of $O(n^2)$, therefore the number of FLOPs is $2n^2$.
    \item The decryption process starts with decoding the ciphertext $y$ applying the permutation $y' = y\cdot P^{-1}$, and using the efficient decoding algorithm for the Goppa code to correct errors in $y'$ and recovering the codeword $c'$ by computing $m'=c'\cdot G^{-1}$, where $G^{-1}$ is the inverse of the generator matrix $G$. Then apply the inverse of the scrambling matrix $m=m'\cdot S^{-1}$, where $m$ is the decrypted message. This also has a complexity of $O(n^2)$ for efficient decrypting algorithms. 
\end{itemize}
% Although McEliece is efficient for encryption, its key generation process is computationally expensive due to the $O(n^3)$ complexity, and it suffers from large key sizes (hundreds of kilobytes to megabytes). Despite these drawbacks, McEliece remains a viable post-quantum candidate due to its long-standing security assumptions based on coding theory.

\subsection{Comparison of Complexities}
CRYSTALS-Kyber and McEliece differ significantly in their mathematical foundations and computational complexities. Kyber, based on lattice problems, is more efficient in terms of key generation and decryption, with smaller key sizes and lower computational overhead. This makes it particularly suitable for constrained environments and applications where key size and computational efficiency are critical. On the other hand, McEliece, based on coding theory, is efficient for encryption but suffers from large key sizes and higher key generation complexity. 
% Although McEliece has a long history of security assumptions, its practical implementation is often hindered by large key sizes. Both cryptosystems have their trade-offs, and the choice between them depends on the specific application requirements, such as key size constraints and computational resources.
Based on Table~\ref{tab:complexity_comparison}, McEliece has a higher complexity compared to Kyber by key generation. This is because McEliece involves matrix-matrix multiplications, which are more expensive than the matrix-vector operations in Kyber. In the encryption process, McEliece is slightly more efficient compared to Kyber, as it only requires matrix-vector multiplication. Kyber is more efficient than McEliece for decryption, as it requires only inner product operations. In terms of parameter sizes, Kyber typically uses smaller parameters, like $n = 256, k=2,3,4$, while McEliece uses larger parameters, such as $n=1024$. This makes Kyber more efficient in practice for key sizes and computational overhead. 

\begin{table}[h]
    \centering
    \caption{Complexity Comparison Between CRYSTALS-Kyber and McEliece}
    \label{tab:complexity_comparison}

    \begin{tabular}{lll}
        \hline
        \textbf{Operation} & \textbf{CRYSTALS-Kyber} & \textbf{McEliece} \\ \hline 
        Key Generation     & $O(k^2 n)$                          & $O(n^3)$                    \\ 
        Encryption         & $O(k^2 n)$                          & $O(n^2)$                    \\ 
        Decryption         & $O(kn)$                             & $O(n^2)$                    \\ \hline
    \end{tabular}

\end{table}

\section{Numerical Analysis Comparison}

The numerical analysis is conducted employing the parameters specified in Table~\ref{tab:kyber_security_levels} and Table~\ref{tab:mceliece_security_levels}. Kyber is characterized by three distinct security levels, each associated with a specific set of parameters. Similarly, McEliece is characterized by a variety of parameter sets that are determined based on the code length and the error-correcting capability.

\begin{table}[h]
    \centering
    \caption{Kyber Security Levels and Parameters}
    \label{tab:kyber_security_levels}
    \begin{tabular}{lcccc}
    \hline
    \textbf{Security} & \textbf{Module} & \textbf{Polynomial} & \textbf{Key Size } & \textbf{Ciphertext} \\ 
    \textbf{Level} & \textbf{Rank $(k)$} & \textbf{Degree $(n)$} & \textbf{(Bytes)} & \textbf{Size (Bytes)}
    \\ \hline
    Kyber512   & 2 & 256 & 800  & 768  \\ \hline
    Kyber768   & 3 & 256 & 1184 & 1088 \\ \hline
    Kyber1024  & 4 & 256 & 1568 & 1568 \\ \hline
    \end{tabular}
\end{table}

\begin{table}[h]
    \centering 
    \caption{McEliece Security Levels and Parameters}
    \label{tab:mceliece_security_levels}
    \resizebox{0.49\textwidth}{!}{%
    \begin{tabular}{lcccc}
    \hline
         \textbf{Security } & \textbf{Code } & \textbf{Error-Correcting} & \textbf{Key Size} & \textbf{Ciphertext} \\
        \textbf{Level} & \textbf{Length ($n$)} & \textbf{Capability ($t$)} & \textbf{(Bytes)} & \textbf{Size (Bytes)} \\ 
        
        \hline

        348864 & 3488 & 64 & 261,120 & 128 \\ \hline
        460896 & 4608 & 96 & 524,160 & 188 \\  \hline
        6688128 & 6688 & 128 & 1,044,480 & 240 \\ \hline

    \end{tabular}
    }
\end{table}

\begin{figure}[htbp]
    \centering
    \resizebox{0.5\textwidth}{!}{%
    \begin{subfigure}{0.4\textwidth}
        \begin{tikzpicture}
            \begin{axis}[
                ybar,
                bar width=0.8cm,
                enlarge x limits=0.3,
                legend style={at={(0.35,0.95)}, anchor=north,legend columns=-1},
                ylabel={Key Size (Bytes)},
                symbolic x coords={512, 768, 1024},
                xtick=data,
                nodes near coords,
                nodes near coords align={vertical},
                ymin=0,
                ymax=2000,
                width=0.95\textwidth,
                height=0.9\textwidth,
                ]
                \addplot[fill=blue!30] coordinates {(512,800) (768,1184) (1024,1568)};
                % \addplot coordinates {(McEliece-348864,261120) (McEliece-460896,524160) (McEliece-6688128,1044480)};
                \legend{CRYSTALS-Kyber}
            \end{axis}
        \end{tikzpicture}
        \caption{Kyber}
        \label{fig:k}
    \end{subfigure}
    \begin{subfigure}{0.4\textwidth}
        \begin{tikzpicture}
            \begin{axis}[
                ybar,
                bar width=0.8cm,
                enlarge x limits=0.3,
                legend style={at={(0.25,0.95)}, anchor=north,legend columns=-1},
                ylabel={Key Size (Bytes)},
                symbolic x coords={348864, 460896, 6688128},
                xtick=data,
                nodes near coords,
                nodes near coords align={vertical},
                ymin=0,
                ymax=1200000,
                width=0.95\textwidth,
                height=0.9\textwidth,
                ]
                % \addplot coordinates {(Kyber512,800) (Kyber768,1184) (Kyber1024,1568)};
                \addplot[fill=red!40] coordinates {(348864,261120) (460896,524160) (6688128,1044480)};
                \legend{McEliece}
            \end{axis}
        \end{tikzpicture}
        \caption{McEliece}
        \label{fig:dfit}
    \end{subfigure}
    }
    \caption{Key Size}
    \label{fig_key_size}
\end{figure}
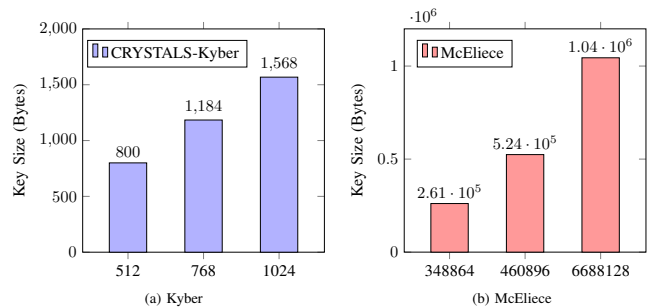

Fig.~\ref{fig_key_size} compares the key sizes of CRYSTALS-Kyber and McEliece across different security levels (128-bit, 192-bit, 256-bit). The key size is a critical metric as it influences both the storage requirements and transmission overhead within cryptographic systems. CRYSTALS-Kyber has significantly smaller key sizes compared to McEliece. For example, Kyber512 (128-bit security) has a key size of 800 bytes, while McEliece-348864 (128-bit security) has a key size of 261,120 bytes. As the security level increases, the key sizes for both grow. However, the key sizes of McEliece continue to be several orders of magnitude greater than those of Kyber. Therefore, Kyber is more suitable for applications with limited storage or bandwidth, such as IoT devices or mobile communication.  

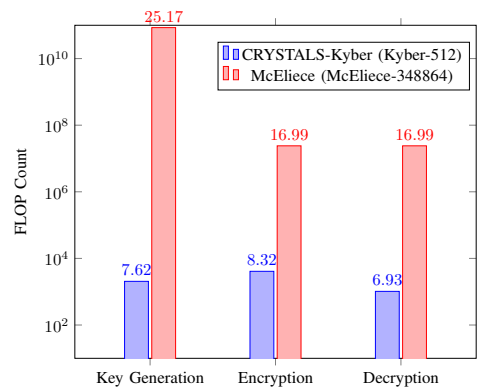
\begin{figure}[htbp]
\centering
    \resizebox{0.35\textwidth}{!}{%
        \begin{tikzpicture}
            \begin{axis}[
                ybar,
                bar width=0.5cm,
                enlarge x limits=0.3,
                legend style={at={(0.67, 0.95)}, anchor=north,transpose legend},
                ylabel={FLOP Count},
                symbolic x coords={Key Generation, Encryption, Decryption},
                xtick=data,
                nodes near coords,
                nodes near coords align={vertical},
                ymin=1e1, % Start y-axis at 10^3
                ymax=1e11, % End y-axis at 10^11
                ymode=log, % Use logarithmic scale for y-axis
                % log ticks with fixed point, % Display log ticks as regular numbers
                ]
                \addplot coordinates {(Key Generation,2048) (Encryption,4096) (Decryption,1024)};
                \addplot coordinates {(Key Generation,8.5e10) (Encryption,2.4e7) (Decryption,2.4e7)};
                \legend{CRYSTALS-Kyber (Kyber-512), McEliece (McEliece-348864)}
            \end{axis}
        \end{tikzpicture}
        }
    \caption{FLOP Count}
    \label{fig_flop_count}
\end{figure}

Fig.~\ref{fig_flop_count} compares the computational complexity of CRYSTALS-Kyber and McEliece in terms of FLOP counts for key generation, encryption, and decryption. FLOP counts provide a measure of the computational effort required for each operation. CRYSTALS-Kyber has much lower FLOP counts for all operations compared to McEliece. For example, Kyber-512 requires 2048 FLOPs for key generation, while McEliece-348864 requires $8.5 × 10^{10}$ FLOPs. Encryption and decryption in Kyber are also significantly faster, with FLOP counts in the thousands, compared to FLOP counts of McEliece in the millions or billions. Kyber is more efficient in terms of computational resources, making it better suited for resource-constrained environments or real-time applications.

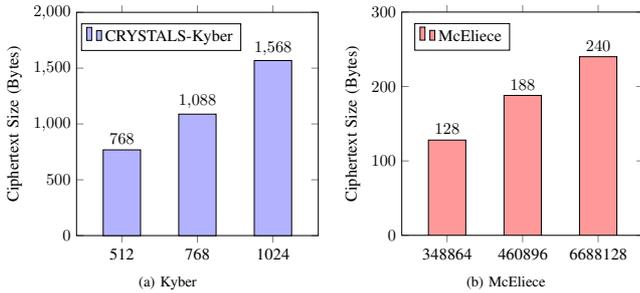
\begin{figure}[htbp]
    \resizebox{0.5\textwidth}{!}{%
    \begin{subfigure}{0.4\textwidth}
        \begin{tikzpicture}
            \begin{axis}[
                ybar,
                bar width=0.8cm,
                enlarge x limits=0.3,
                legend style={at={(0.35,0.95)}, anchor=north,legend columns=-1},
                ylabel={Ciphertext Size (Bytes)},
                symbolic x coords={512, 768, 1024},
                xtick=data,
                nodes near coords,
                nodes near coords align={vertical},
                ymin=0,
                ymax=2000,
                width=0.95\textwidth,
                height=0.9\textwidth,
                ]
                \addplot[fill=blue!30] coordinates {(512,768) (768,1088) (1024,1568)};
                % \addplot coordinates {(McEliece-348864,128) (McEliece-460896,188) (McEliece-6688128,240)};
                \legend{CRYSTALS-Kyber}
            \end{axis}
        \end{tikzpicture}
        \caption{Kyber}
        \label{fig:k}
    \end{subfigure}
    \begin{subfigure}{0.4\textwidth}
        \begin{tikzpicture}
            \begin{axis}[
                ybar,
                bar width=0.8cm,
                enlarge x limits=0.3,
                legend style={at={(0.25,0.95)}, anchor=north,legend columns=-1},
                ylabel={Ciphertext Size (Bytes)},
                symbolic x coords={348864, 460896, 6688128},
                xtick=data,
                nodes near coords,
                nodes near coords align={vertical},
                ymin=0,
                ymax=300,
                width=0.95\textwidth,
                height=0.9\textwidth,
                ]
                %\addplot coordinates {(Kyber512,768) (Kyber768,1088) (Kyber1024,1568)};
                \addplot[fill=red!40] coordinates {(348864,128) (460896,188) (6688128,240)};
                \legend{McEliece}
            \end{axis}
        \end{tikzpicture}
        \caption{McEliece}
        \label{fig:dfit}
    \end{subfigure}
    }
    \caption{Ciphertext Size}
    \label{fig_ciph_size}
\end{figure}

Fig.~\ref{fig_ciph_size} compares the ciphertext sizes of CRYSTALS-Kyber and McEliece at different security levels. The size of the cryptotext is important because it affects the amount of data that must be transmitted during encryption. CRYSTALS-Kyber has larger ciphertext sizes compared to McEliece. For example, Kyber512 has a ciphertext size of 768 bytes, while McEliece-348864 has a ciphertext size of 128 bytes. However, the difference in ciphertext sizes is much smaller than the difference in key sizes. Although McEliece has smaller ciphertexts, its large key sizes and high computational complexity make it less practical for many applications. The marginally larger ciphertexts of Kyber are counterbalanced by its reduced key sizes and diminished computational overhead.

% \section{Future Research Directions}
% \begin{itemize}
%     \item New Mathematical Techniques: Developing new mathematical techniques for efficient computation can help in reducing the hardware complexity of PQC algorithms. Research in this area can lead to more efficient implementations.
%     \item In-Memory Access Optimization: Optimizing in-memory access can help in reducing the memory requirements of PQC algorithms. Techniques like data compression and efficient memory management can be explored.
%     \item Hardware-Software Co-Design: A co-design approach, where both hardware and software components are optimized together, can lead to more efficient implementations of PQC algorithms in embedded systems.
% \end{itemize}

\section{Conclusion}
The paper examines the implementation of post-quantum cryptography (PQC) in embedded systems constrained by computational power, memory, and energy. It categorizes PQC algorithms into families: lattice-based, code-based, hash-based, and multivariate/isogeny-based schemes, each with unique challenges. Lattice-based schemes like CRYSTALS-Kyber require moderate memory and significant hardware for polynomial arithmetic. Code-based schemes like McEliece have large key sizes. Hash-based schemes such as SPHINCS+ have simple computation but large signatures, while multivariate/isogeny-based schemes demand too many resources. Optimization through pipelining, parallelization, and high-level synthesis can improve performance and energy efficiency while balancing security. CRYSTALS-Kyber suits constrained environments for key generation and decryption, unlike McEliece. Future research should explore new techniques and memory improvements to reduce hardware complexity, as quantum computing develops.

\bibliographystyle{IEEEtran}
\bibliography{references}

\end{document}